\documentclass[12pt]{iopart}
\usepackage{graphicx}
\usepackage{amsmath}

\begin{document}

\title[.]{A study of the coupled dynamics of asymmetric absorbing clusters in a photophoretic trap}

\author{Anita Pahi$^1$, Shuvojit paul$^2$ and Ayan Banerjee$^1$}

\address{$^1$Indian Institute of Science Education and Research, Kolkata 741246,$^2$Kandi Raj College, Kandi, Murshidabad 742161, India}
\ead{\mailto{ayan@iiserkol.ac.in}}

\vspace{10pt}

\begin{abstract}
We report a study on the dynamics of absorbing asymmetric carbon clusters trapped by a loosely focused Gaussian beam using photophoretic force. At high laser powers, all the trapped clusters display rotation coupled with oscillation along the axial direction, with a majority spinning about a body fixed axis, while the rest display dual spin as well as orbital motion about a fixed point in space. The spinning and orbiting frequency is inversely proportional to the amplitude of the axial oscillation - with one growing at the expense of the other. Further, the frequencies of these rotations are not proportional to the laser power, but to the trap stiffnesses inferred from the corresponding natural frequencies. The clusters also stop rotating below a certain laser power, and execute random thermal fluctuations.  Our work suggests that the dynamics of clusters trapped with photophoretic force are largely dependent on the cluster size and morphology, which could, in principle, be tuned to obtain various motional responses, and help in the design of rotating micromachines in air.
\end{abstract}

%
%
%
%
%

\section{Introduction}
Photophoretic forces have emerged as a robust enabler of the trapping of absorbing clusters, both in air and water \cite{shvedov2010giant,shvedov2009optical,paul2022optothermal}, using even loosely focused fundamental Gaussian beams \cite{bera2016simultaneous,sil2020study}. 
Their efficacy and ease of implementation has encouraged researchers to develop a wide range of applications, including 3D volumetric displays \cite{smalley2018photophoretic}, massive cluster manipulation \cite{shvedov2010giant}, and cluster sorting through controllably changing force direction  \cite{helmbrecht2007photophoretic}. However, the understanding of this force and the dynamics of trapped clusters using this force is yet far from complete. A concept rather prevalent has been that structured beams\cite{shvedov2009optical} are essential for photophoretic trapping - something that has been disproven by numerous reports of trapping using simple Gaussian beams{\cite{bera2016simultaneous,sil2020study,he2019investigation}}. Nevertheless, larger temperature gradients may be applied to absorbing clusters by structured beams due to the existence of intensity minimas in them, leading to stronger forces \cite{shvedov2009optical,sil2024ultrastable,pahi2023comparison}. 
However Gaussian beams are much simpler to model theoretically, so that the understanding of the dynamics of photophoretically trapped clusters using Gaussian beams should be a crucial first step. 

Now, photophoretic forces are of two types - the $\Delta T$ force, which depends on the temperature gradient across the absorbing cluster, and the $\Delta \alpha$ force - which depends on the morphology or accommodation coefficient of the cluster \cite{horvath2014photophoresis}. 
Indeed, while trapping in the axial direction for clusters falling against gravity may be explained as the phenomenon of levitation (the light intensity being adequate to generate enough photophoretic force to balance gravity), trapping in the radial direction (perpendicular to the beam propagation) is more complex since it requires the existence of a restoring force - which may be generated due to the inhomogeneous intensity perceived by the cluster - and which may lead to complex trajectories and dynamics inside the beam profile. Wurm et al.\cite{wurm2008experiments} were the first to talk such dynamics - viz. the rotation of graphite clusters of size several tens of $\mu$m, photophoretically trapped in a loosely focused Gaussian beam. However, systematic analysis of the dynamics of these trapped clusters is lacking, except for a few instances \cite{lin2014optical,chen2018temporal,liu2014manipulation}.\cite{ex1}. In addition, the possibility of applying such forces in the development of optically controlled micromachines in air, where the machines may be subjected to both rotational and orbital motion, and even stopped controllably - has not been explored - to the best of our knowledge.

In this paper, we demonstrate such controllable rotational motion on optically trapped carbon clusters of size around $\sim 20~\mu$m (measured in the direction transverse to the laser beam), which is comparable to the laser focal spot size. The clusters demonstrate both spin and orbital motion, and may even be stopped at low laser power. Interestingly, we observe that the dynamics of the clusters is highly dependent on morphology and alignment with the laser beam, with a majority of the trapped clusters spinning about a body fixed axis, while the rest both spin and orbit about a direction depending on the shape of the cluster. Along with these rotations, the trapped clusters also oscillate along the beam propagation direction, and the amplitudes of these oscillations appear to be inversely proportional to the frequencies of rotations of these trapped clusters. For all clusters, the stiffness - as expected - is always proportional to the spinning frequency of the cluster. Below a certain laser power, the  clusters remain trapped but stop rotating, undergoing only thermal fluctuations. Our results thus demonstrate the viability of utilizing photophoretic forces for developing micromachines in air capable of complex rotational motion, which can also be stopped controllably.  We first describe the experimental setup and later analyse the results using various theoretical and experimental tools.

\section{Experimental set up}

\begin{figure}[htbp]
\includegraphics[width = 1 \textwidth]{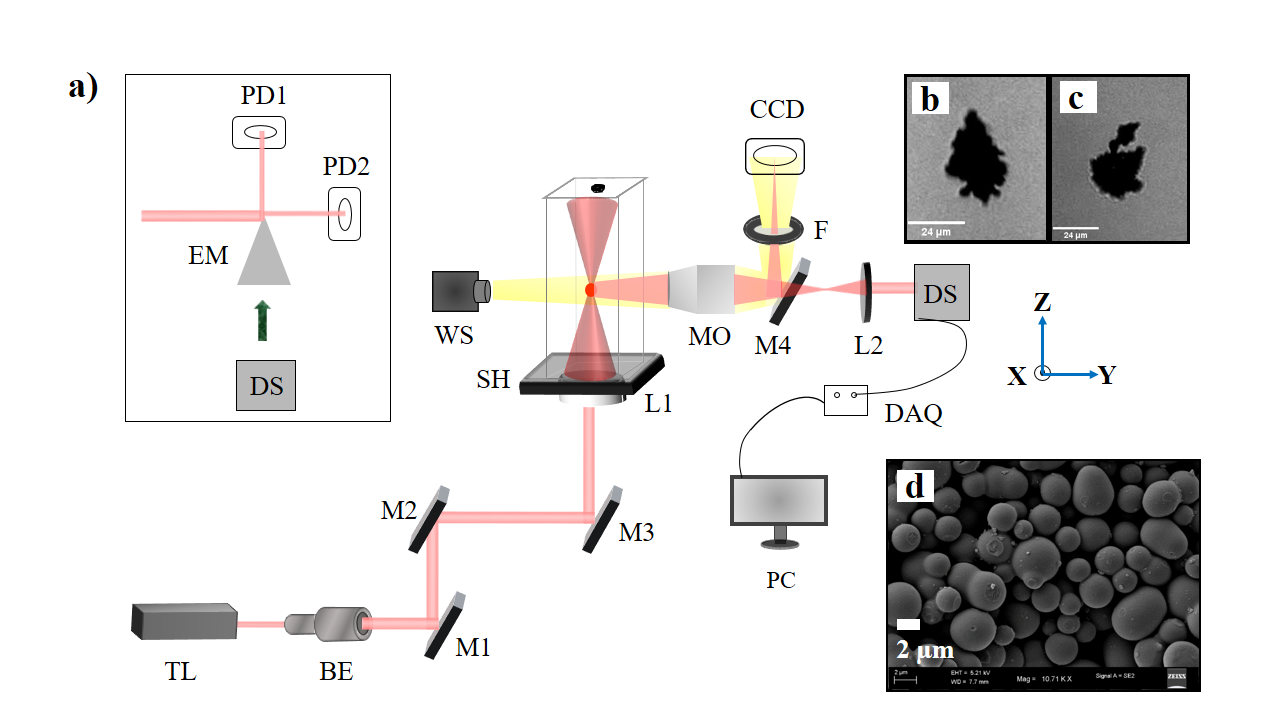}
\caption{(a) Schematic of the experimental set up. Labels denote the following, TL : Trapping laser, BE : Beam expander, M : Mirror, L : Lens, SH : Sample holder, MO: Microscope objective, WS: White light source, F : Filter, CCD : Camera , DS : Detection system, DAQ: Data acquisition card, PC: Computer, PD1 and PD2 : Photodiode 1 and photodiode 2, EM : Edge mirror, (b) and (c) shows two different types of trapped clusters, (d) SEM image of polydisperse carbon microspheres}
\label{schematic}
\end{figure}

\begin{figure}[htbp]
\includegraphics[width = 1 \textwidth]{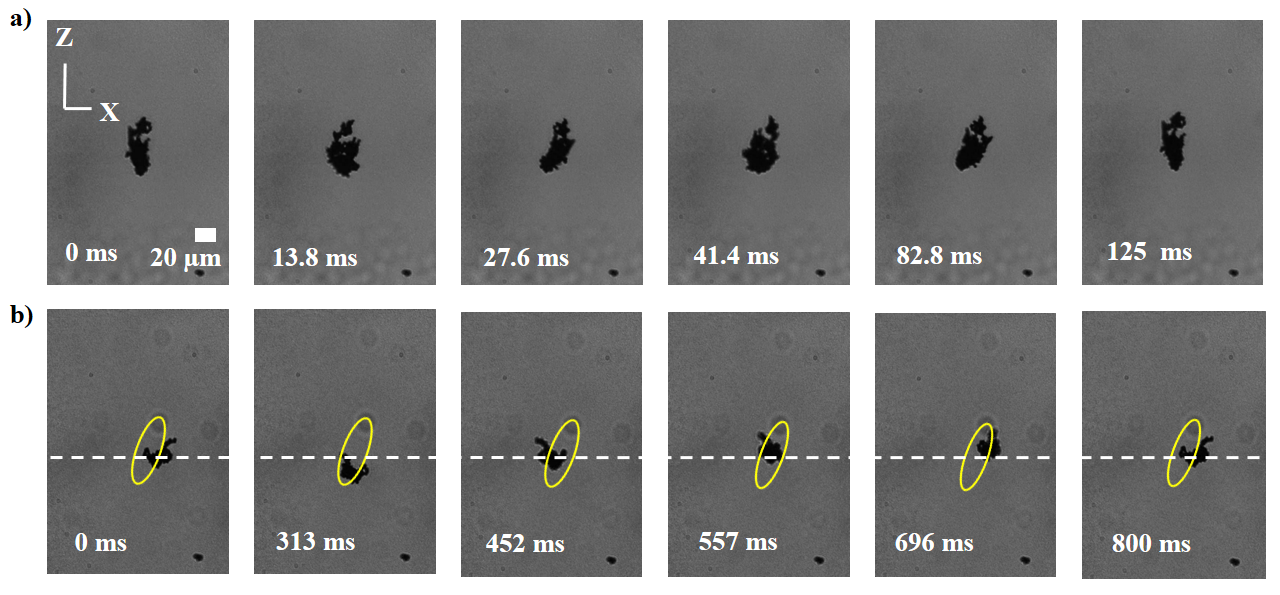}
\caption{(a) Time-lapsed images of a cluster spinning about a body-fixed axis, as is evident from changes in the cluster's orientation (frames extracted from \textit{supplementary video} 1). The cluster finished one full rotation between the first and last frames as shown from its orientation in the first and sixth frames. (b) Time-lapsed images of a cluster both spinning and orbiting at different times showing one full rotation in an orbit marked by a yellow line (frames extracted from the \textit{supplementary video} 2). The spinning about a body-fixed axis is evident in changes in the cluster's orientation, while orbiting is clear from different positions of the cluster on the orbit. The white line drawn acts as a reference, indicating the change in the z position of the cluster at different frames.}
\label{exptpics}
\end{figure}

We used a $25$ mm focal length lens to loosely focus a Gaussian laser beam of wavelength $\lambda$ = 640 nm from a standard semiconductor laser (Oxius) to trap a cluster of poly-dispersed carbon microspheres (Sigma Aldrich) of size between 2-13 $\mu m$.  One schematic of the setup is shown in the Fig.~\ref{schematic}(a). From the image taken through a SEM, the average size of the carbon microspheres is 5 $\mu m$. The microspheres form highly asymmetric clusters due to adhesive forces between carbon clusters, and the measured average size of the trapped clusters is 20 $\mu$m. Histograms displaying the size distribution of both the individual microspheres and the clusters is provided in Fig.~1 of the \textit{Supplementary Information}. A typical picture of such clusters is shown in Fig.~\ref{schematic}(b)-(c). We trap one of these clusters inside a homemade cuboidal glass sample chamber, where microspheres are placed upside down on the top glass lid of the sample chamber and are introduced into the sample chamber by a physical perturbation of the lid. They are seen falling in clusters vertically downwards, which we trap using the laser propagating along the vertically upward ($z$) direction, with the beam waist being approximately at the center of the sample chamber (see Fig.~\ref{schematic}(a)). The size of the beam waist is measured to be 19.6 $ \pm 0.5~\mu m$ using the Knife edge technique \cite{sil2017dual}. As some of the clusters fall into the beam path, they experience the photophoretic force and, depending on cluster size, morphology and laser intensity, get spatially confined in all three dimensions. Importantly, different clusters get trapped at different axial positions depending on their mass, morphology and size. However, we observe that almost all the clusters get trapped at regions slightly above the focus. 

Initially, we used around $80$ mW laser power to trap a cluster, and then decreased the power up to $10$ mW to study the resultant effect on the trapped cluster dynamics. Scattered light from the trapped cluster in a direction perpendicular to the trapping beam is collected using a 10X, 0.25 NA dry microscope objective (Olympus). After the 10X objective, a mirror is placed on a flip-mount, which can direct the scattered light to the CCD. A 600 nm low pass filter is placed before the CCD to cut off the trapping beam and to image the cluster using white light emanating from a mercury lamp source. When the flip-mount is removed from the beam path, the scattered light reaches a balanced detector system to measure the position fluctuations of the trapped cluster. A balanced detector system is a combination of two photodiodes that amplifies the difference in the signals of two, and rejects common mode noise  \cite{bera2017fast}. Data is collected using a data acquisition card (NI-DAQ) each for $50$s at a sampling frequency of $20$kHz. To measure the size of the cluster, the CCD pixels are first converted into $\mu$m using a micrometer scale, and later, it is used to find the width of the trapped cluster, i.e., its maximum extent in the $x$ direction (transverse to the beam propagation direction). To measure the position fluctuations of the trapped cluster, we remove the flip mirror and use the balanced detection system to record the corresponding time series. By putting the flip mirror back, we can also record the dynamics of the trapped cluster using the CCD camera. 

At higher laser powers, the trapped cluster execute two different kinds of motion: a) rotation about a body-fixed axis of the cluster (the axis being parallel to the direction of the trapping beam as displayed in the \textit{Supplementary Videos} 1 and 2, note that we shall henceforth refer to this rotation as spin), b) spin coupled with an orbital rotation in a direction depending on the shape of the cluster (\textit{see Supplementary video} 3). In the Fig.~\ref{exptpics}(a), we have shown the snapshots of a spinning cluster at $60$ mW laser power (extracted from \textit{supplementary video} 1), where the spin is clear from the change of the orientation of the cluster. On the other hand, in the Fig.~\ref{exptpics}(b), we have shown the same for the orbiting cluster at the same laser power. It is apparent that the cluster orbits about a point in space (which we can refer to as the trapping center) with an additional spin. However, for the orbiting cluster, the spinning frequency appears exactly the same as the orbiting frequency when the first and the last frames Fig.~\ref{exptpics}(b) are compared. Further, when the laser power increases, we observe that the orbit diameter also increases, with a concomitant decrease in orbiting (or spinning) frequency. 

For all clusters (undergoing both types of rotations), we also observe an oscillation along the beam propagation direction (i.e., along the $z$ axis). This change of the cluster position along the $z$ direction (axial) is apparent in the supplementary videos (\textit{see supplementary videos} 1 to 5). We have observed such axial oscillations in earlier work \cite{sil2024ultrastable}, and attribute this to an interplay between photophoretic forces and gravity, as we explain later in the paper. 

The clusters usually get trapped at a height slightly above the focal plane; however, they move towards the focus as the laser power is decreased. Interestingly, at a laser power of $\sim 10$ mW, when the clusters come very close to the focal plane, we observe both the spinning and orbiting to cease, and the trapped clusters display a jittery motion which we believe to be due to thermal fluctuations. With further decrease in power, the clusters leave the trap. 

\section{Data Analysis}

Now, the one-dimensional (along the $x$ direction) dynamics of a non-rotating cluster of mass $m$ trapped in air (low viscous medium) with trap stiffness $k$ can be described by the following Langevin equation:
\begin{equation}
    \ddot{x} + g\dot{x} + \Omega^{2}x = \Lambda\zeta(t).
    \label{eq1}
\end{equation}
Considering the friction coefficient along the observation direction being $\gamma$, $g = \frac{\gamma}{m}$. $\Omega = \sqrt{\frac{k}{m}}$ is the natural frequency of the Brownian harmonic oscillator, $\Lambda$ is the strength of the thermal noise, and $\zeta$ is the Gaussian distributed, delta-correlated random noise representing the thermal fluctuations. Solving the above Eq.~\eqref{eq1} for the power spectral density (PSD), one can obtain
\begin{equation}
    S(\omega) = A\frac{\Omega^2g}{(\Omega^2 - \omega^2)^2 + \omega^2g^2},
    \label{eq2psd}
\end{equation}
where, $S$ represents the PSD and $\omega$ represents the angular frequency. This PSD, however, should be different for the rotating clusters. As any rotation of a trapped cluster observed from one-dimension appears like oscillations, the corresponding peaks should be added to the related PSDs. We have shown the PSDs corresponding to the spinning and rotating clusters that contain peaks due to rotations with harmonics up to 40-50 orders in the Figs.~\ref{PSD}(a)-(b). Notably, the $x$ directional trajectory of one of the rotating clusters (see Fig. 2 of the \textit{SI}) shows that the rotation appears to be rather complex, which may be due to the morphology of the cluster. Therefore, higher-order frequencies should appear in the PSD along with the fundamental rotation frequency. We determine the spin or orbital frequency of the rotating clusters from the corresponding fundamental peak. However, to fit the Eq.~\eqref{eq2psd} to the PSDs related to the rotating clusters and to infer parameters (such as natural frequency $\Omega$), we first run the PSD through a peak removal algorithm in MATLAB, and then block each PSD with $30$ points in a bin. Two representative results are shown in Figs.~\ref{PSD}(b) and (d), along with the corresponding fits.
\begin{figure}[htbp]
\includegraphics[width=1\textwidth]{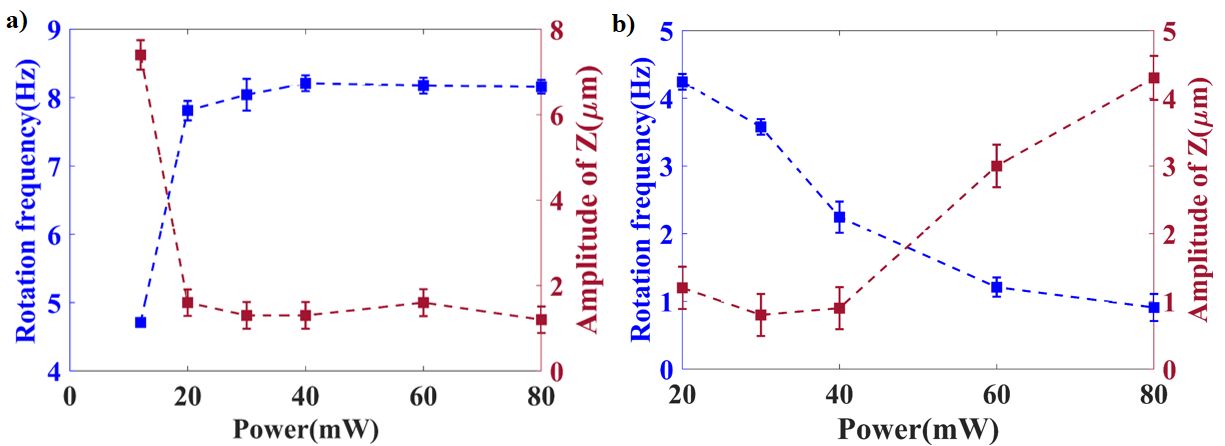}
\caption{Values of spinning frequency and amplitude of Z oscillation as function of laser power. (a) Represent spinning particles, and (b) represents orbiting particles.}
\label{amp_power}
\end{figure}

\begin{figure}[htbp]
\includegraphics[width=1\textwidth]{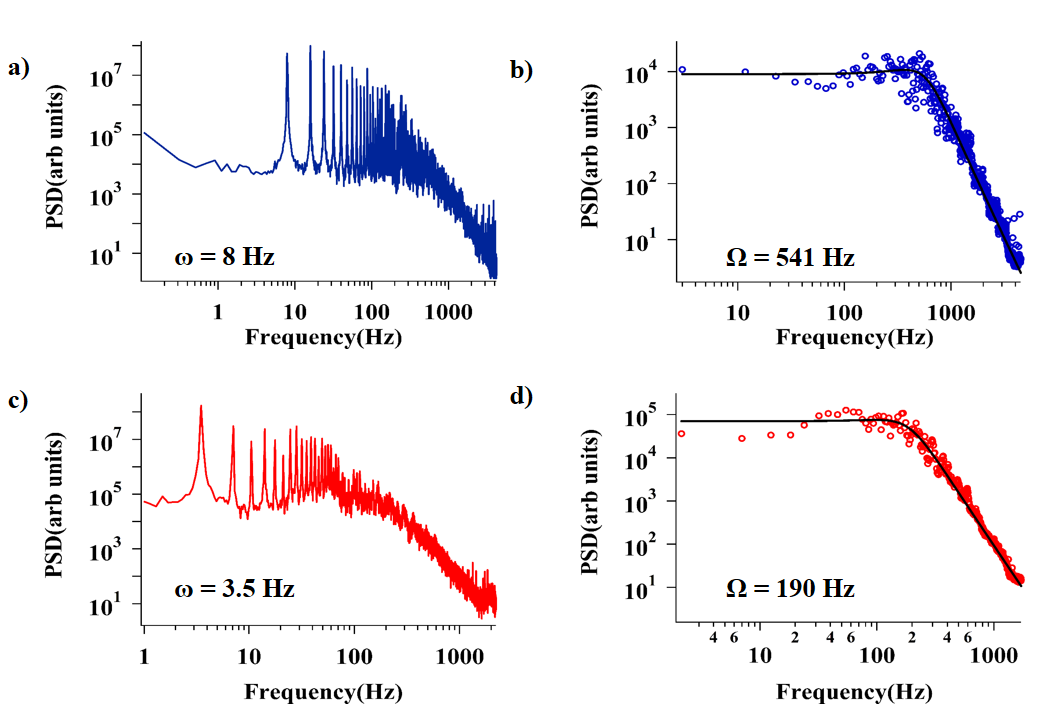}
\caption{Power spectral density from the position fluctuations of the trapped clusters. (a) PSD of the spinning cluster at 30 mW . (b) Fit of Eq.~\ref{eq2psd} to the peak removed PSD of the spinning cluster. (c) PSD of the orbiting cluster at 30 mW.  (d) Fit of Eq.~\ref{eq2psd} to the peak removed PSD of the orbiting particle. Inset values depict frequencies of fundamental rotation peaks for (a) and (c), and resonance frequencies for (b) and (d).}
\label{PSD}
\end{figure}

\section{Discussion}
\begin{figure}[htbp]
\includegraphics[width=1\textwidth]{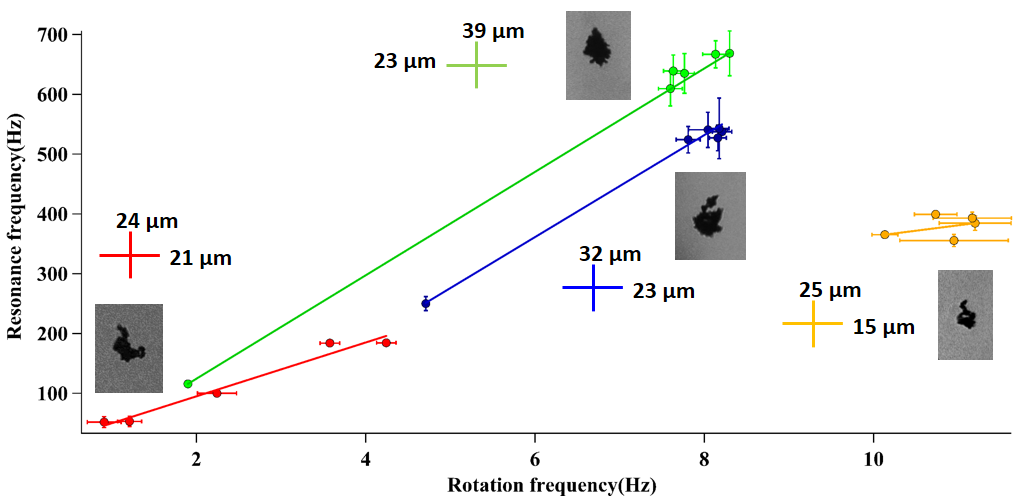}
\caption{Resonance frequency versus rotation frequency of various clusters (different colours) with linear fits. Image of the particular cluster along with its size is displayed next to the data.}
\label{spinstiffness}
\end{figure}

As mentioned earlier, one of the most interesting results we report here is the behaviour of the two categories of clusters we have studied, where one category appears to only spin, while the other both spin and undergo orbital rotation. We have studied a total number of 15 clusters, out of which $\sim 80~\%$ displayed only spin, while the rest ($\sim 20~\%$) exhibited both spin and orbital motion. There were several key quantitative takeaways from the data: 1) For only spinning clusters, the frequency initially increases with laser power, and after $\sim 30$ mW, it becomes almost constant, as shown in Fig.~\ref{amp_power}(a) (experimental data depicted in blue diamonds). Interestingly, the amplitude of z oscillation shows an inversely proportional relation and decays with power (some of the trajectories in the $z$ direction are shown in the \textit{SI}) before becoming constant at a laser power of around $30$ mW. The amplitude of the $z$ oscillation with laser power is plotted in the right axis of the Fig.~\ref{amp_power}(a). For clusters executing both spin and orbital motion, the spinning frequency (which is the same as the orbital frequency) clearly reduces gradually with increasing laser power, as demonstrated in Fig.~\ref{amp_power}(a), with an associated increase in orbital radius. For these clusters, as expected, the $z$ motion is complex compared to the spinning clusters, as the cluster explores more regions of the beam, and the change of the cluster orientation is comparatively large. Even here we observe an approximate inverse trend of the amplitude of the $z$ oscillation with the cluster spin frequency (see the \textit{Supplementary Information} for the corresponding trajectories). (2) The spin frequency is a clear indicator of the trap stiffness. In all cases, we observe a direct correlation between the spin frequency and the trap stiffness signified by the natural frequency, which we explicitly display in Fig.~\ref{spinstiffness} with four different clusters (different colors represent different clusters), where we also provide robust linear fits to each set of data. We have also determined sizes of the trapped clusters along x and y direction, by approximating them as ellipsoids, and measuring the major and minor axes (shown in Fig.~\ref{spinstiffness} as coloured crosses), the horizontal line depicting the $x$, and the vertical line the $y$ measurement. There does not seem to be a direct correlation between the natural frequency and size, which may suggest that cluster morphology plays an important role in determining the resonance frequency. We have also attached the images of the trapped clusters  in Fig.~\ref{spinstiffness}, which clearly indicate their diverse morphology. (3) All the rotating trapped clusters come very close to the focal point below a certain laser power $\sim 10$ mW and stop rotating, but fluctuate randomly due to thermal effects. To the best of our knowledge, this phenomenon has not been reported before, and we speculate it happens due to the size of the clusters being similar to the beam spot size.

We now attempt to understand these results qualitatively. Orbital motion of particles trapped by photophoretic forces has been observed earlier \cite{lin2014optical}, and has been attributed to the transverse component of the photophoretic $\Delta~\alpha$ force generating a torque due to the inherent particle asymmetry, which causes the body force and the gravitational force to act on different points on the particle surface.  This is what we observe in our case as well, with the clusters being extremely asymmetric in general. However, we also observe a spinning motion in our trapped clusters along with the orbital motion, and for certain types of particles, we only observe spinning. Interestingly, for these, the mass distribution around an axis passing through the trap center appears more symmetric than for the orbiting clusters (\textit{see the Supplementary videos}), which is consistent with our understanding. Note that spinning motion of asymmetric particles trapped in optical tweezers in water due to imbalance in radiation pressure forces generated due to asymmetric scattering of light from them is known \cite{hirugashi1999pre}. In our case, a theoretical study needs to be conducted, possibly using the T-matrix method \cite{Michael2004JQuant} to determine the Maxwell stress tensor and thereby the scattering forces and torque acting upon the clusters trapped in air. Note that torque may also be generated by the $\Delta~\alpha$ force, arising due to the morphology of the cluster.

Further, it appears that clusters that do not orbit experience almost the same trap stiffness when the laser power is increased. This is expected since the cluster mass is unchanged throughout the experiment, so the same laser intensity is required to balance gravity, as we had shown earlier with printer toner particles trapped using a loosely focused Gaussian beam \cite{sil2020study}.  Indeed, as described earlier, we do observe a shift in the clusters' mean axial position towards the beam focus as we reduce laser power - which implies that the clusters tend to seek similar laser intensity regions when the laser power is modified. On the other hand, for clusters that both spin and orbit - the dynamics is largely different due to their morphology. Instead of spin, here, the radius of the orbit increases with power. 

The associated oscillation along the axial direction is due to an interplay between photophoretic forces and gravity \cite{sil2024ultrastable}, with the clusters rising upwards when the former dominates, and thereby reaching a region of low laser intensity where they fall down when gravity dominates. However, with an increase of cluster spin or orbital rotation frequency, the oscillation amplitude decreases. We believe that this is the case since the origin of both phenomena is the same, viz. photophoretic forces, which depend both on laser intensity as well as particle morphology - whose interplay may lead to different dynamics of the clusters  at different laser powers. 

Finally, optical-tweezers operated micromachines have been demonstrated widely in liquids \cite{yinan2022softm, Zhang2021} - mostly water, with the morphology of the machines and laser power being the principal controlling factors driving the motion of the micromachines. However, such studies in air have not been performed to the best of our knowledge - mostly due to the fact that merely trapping particles using radiation pressure forces in air is significantly more difficult compared to that in water, so that controlled manipulation is virtually out of the question. Photophoretic forces, however, make trapping of particles in air much more simpler. Our work, however, clearly shows that such forces may also be used in developing micromachines displaying complex rotational motion in air, with a high degree of control being exerted on the motion of the micromachines by simply changing laser power. While our micromachines display various types of motion including spin and orbital rotation, as well as oscillation, the fact that these motions are coupled inversely, allows for the choice of a particular motional mode, simply by adjusting laser power and micromachine morphology. While in this case we have not attempted to control the latter - our work clearly shows that the more the asymmetry, more is the propensity of the machine to execute complex rotational motion, which - most importantly, can be modified by simply adjusting laser power.

\section{Conclusion}

In conclusion, we trap highly asymmetric carbon micro-clusters under the influence of photophoretic forces and observe two distinct types of dynamics - one where the clusters undergo a spinning motion along a body-fixed axis in the direction of the laser beam and the other where the clusters simultaneously orbit and spin. Both these kinds of clusters also oscillate along the beam propagation direction, which seems to be anti-correlated to the rotational dynamics. The orbital and spinning motions seem to be predominantly dependent on the cluster morphology - since clusters of similar mass indulge in either of the two dynamics. Importantly, the spinning motion is always positively correlated with the trap stiffness, and clusters that only spin experience almost the same trap stiffness when the laser power is increased. This is understandable since clusters with the same mass need the same intensity of light to balance gravitational forces. However, when the laser power is low enough so that the cluster reaches the focus, we observe trapping without rotation too - which happens possibly since the size of the trapped cluster is comparatively larger with respect to the beam spot size. Finally, given the importance of size and shape in producing particular types of rotational motion, particles may thus be engineered appropriately to generate a desired rotational motion, so that new routes of creating micromachines in air may be envisaged.  We plan to work in these areas using manicured particles, as well as adaptive optics-generated complex beam profiles to generate even more intriguing but controllable dynamics of trapped particles in air. 

\section{Data Availability statement}
The data that support the findings of this study are available upon reasonable request from the authors.

\section{Acknowledgements}
This work was supported by the Indian Institute of Science Education and Research, Kolkata, an autonomous research and teaching institute funded by the Ministry of Education, Government of India.

\section{Appendix}
\subsection{Cluster size measurement}
\begin{figure}
    \centering\includegraphics[width=1 \linewidth]{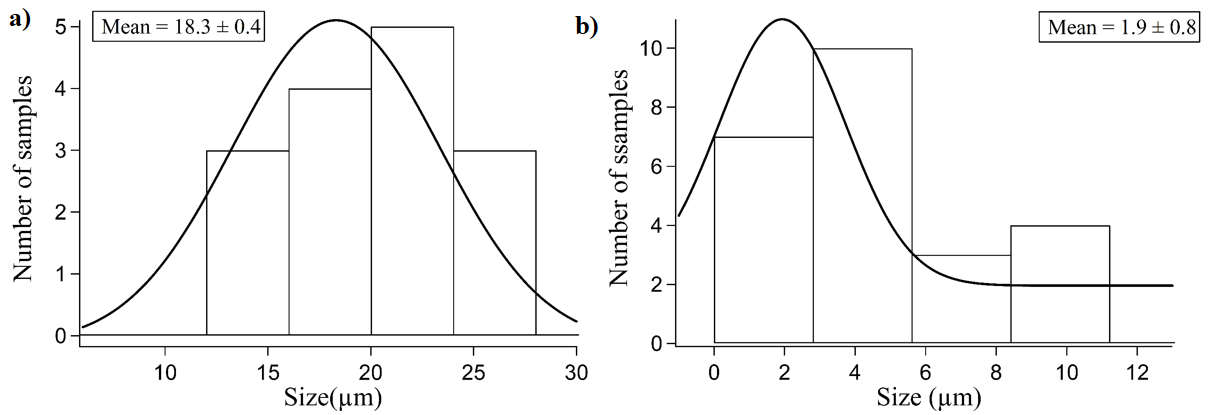}
\caption{Size distribution fitted to a normal distribution (a) size distribution of trapped clusters (b) Size distribution of carbon microspheres where size was measured from a SEM image}
\label{histogram}
\end{figure}

The particles used for trapping are carbon microspheres. We used a SEM to image these microspheres with high resolution. Using the calibration factor provided in the SEM software, we determined the size of the carbon microspheres using ImageJ software, which came out to be from 2 to 11 $\mu$m. Additionally, the size of the trapped clusters was measured using a CCD. The field of view of the CCD was calibrated using a micrometer scale. Both size distributions are shown in Fig.S1 \ref{histogram}. We observe that the distributions are reasonably Gaussian, with the mean size of the trapped clusters being $18.3\pm 0.4 ~\mu m$, while that of the individual particles by SEM imaging is $2\pm 0.8 ~\mu m$.

\subsection{Z position measurement}

\begin{figure}
    \centering\includegraphics[width=1 \linewidth]{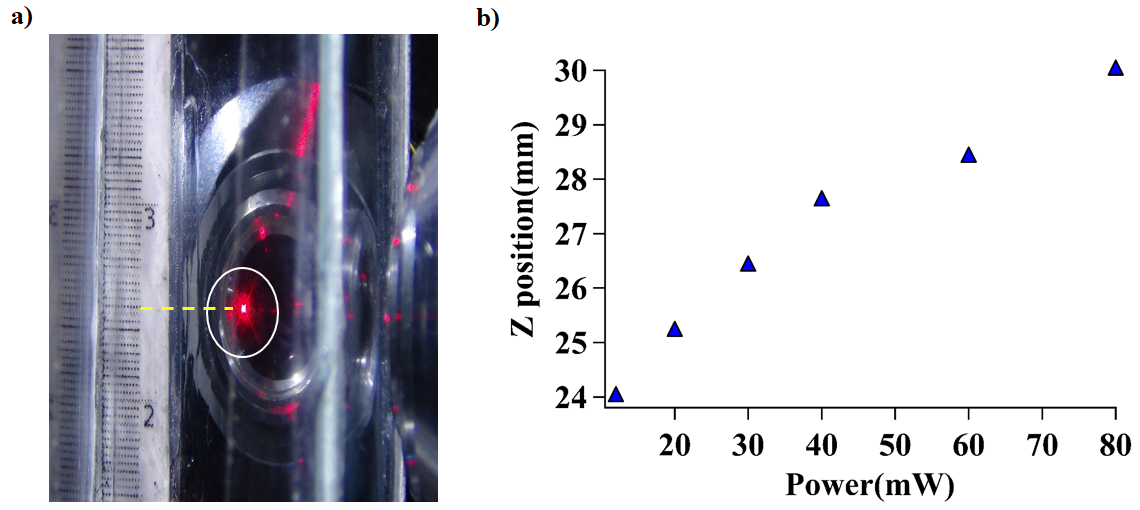}
\caption{Z position measurement (a) Image of the particle inside the sample chamber with the scale attached (b) Z position of one particle as a function of power}
\label{z_position}
\end{figure}

The z position of the trapped clusters was measured using a scale attached to the sample chamber, as shown in Fig.S\ref{z_position}(a). The clusters are initially trapped at a laser power higher than the minimum power required for trapping. Once trapped, we reduce the laser power to observe the variation in their dynamics at different powers. Similar to the behavior observed with toner particles\cite{sil2020study}, these clusters change their axial ($z$) position at different powers (Fig.S\ref{z_position}(b)). They move towards the focal point as the power decreases, seeking the same intensity at various powers. This specific intensity is necessary to balance their mass.

\section{Analysis}
\begin{figure}
   \centering\includegraphics[width=1 \linewidth]{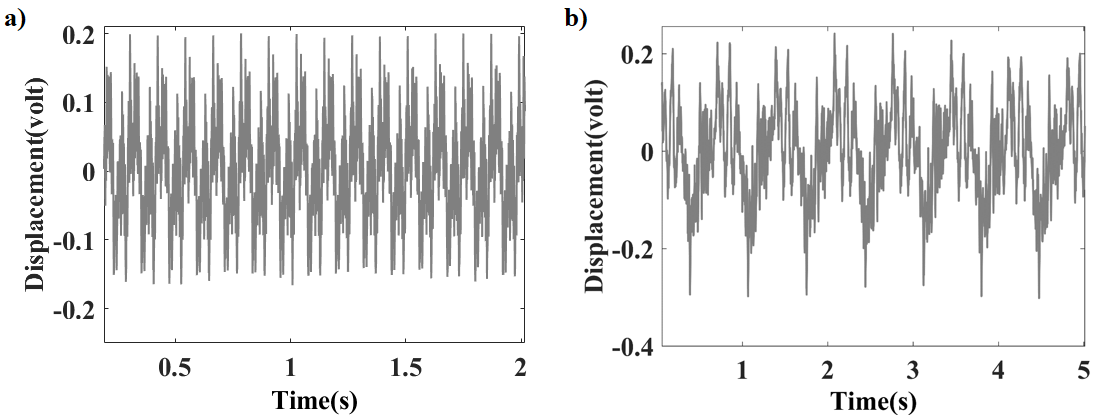}
\caption{Time series of the scattered light from trapped cluster measured using balanced detection. (a) Time series of spinning particle at 60 mW (b) Time series of orbiting particle at 60 mW.}
\label{X Time series}
\end{figure}

\begin{figure}
    \centering\includegraphics[width=1 \linewidth]{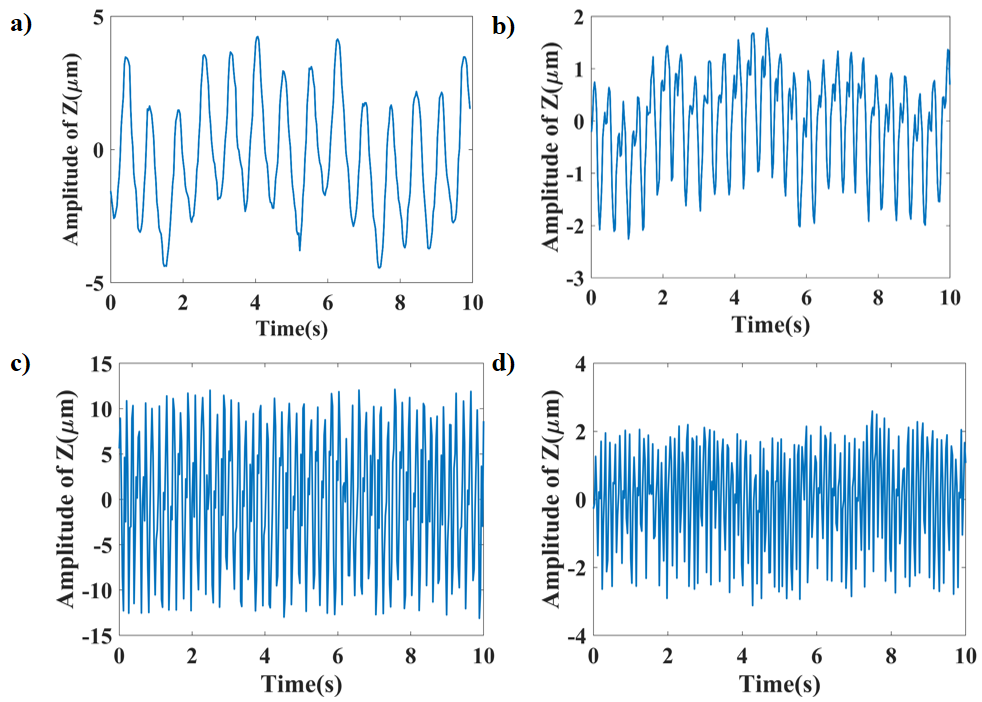}
\caption{Z time series of the trapped cluster from video analysis (a) Z time series of orbiting particle at 60 mW (b) Z time series of orbiting particle at 40 mW. (c) Z time series of spinning particle at 12 mW. (d) Z time series of spinning particle at 20 mW}
\label{Z Time series 20mW}
\end{figure}
We measure the position fluctuations of the trapped clusters using a balanced detection method consisting of two photodiodes. Two representative time series of a cluster that is spinning and orbiting at 2 different powers (60 and 40 mw, respectively) are shown in Fig.S\ref{X Time series}(a) and (b). The time series exhibit a distinctive periodic nature due to the rotation of the cluster along with thermal fluctuations. Additionally, the cluster shape is quite asymmetric, causing the scattered light from the cluster to vary with small change in cluster orientation. These factors contribute to a complex time series. However, the overall periodic form reveals the fundamental rotation frequency, which is also obtained from power spectrum analysis. We also record Videos of the trapped clusters along the XZ plane. By tracking the centroid of the cluster in MATLAB, we obtained the z trajectories of the trapped clusters. Using these trajectories, the amplitude of z oscillation was measured. Trajectories corresponding to both spinning and orbiting clusters has been shown in Fig.S\ref{Z Time series 20mW} (a)-(d),  respectively. 

\section{References}
\bibliography{iopart-num}

\end{document}